# Novel ion-doped mesoporous glasses for bone tissue engineering: study of their structural characteristics influenced by the presence of phosphorous oxide.

Anahí Philippart[1], Natividad Gómez-Cerezo[2,3], Daniel Arcos[2,3], Antonio J. Salinas[2,3], Elena Boccardi[1], MariaVallet-Regi[2,3]*, Aldo R. Boccaccini[1]*

[1.] Institute of Biomaterials, University of Erlangen-Nuremberg, Cauerstraße 6, 91058 Erlangen, Germany

[2.] Departamento de Química Inorgánica y Bioinorgánica. Facultad de Farmacia, Universidad Complutense de Madrid. Instituto de Investigación Sanitaria Hospital 12 de Octubre i+12. Plaza Ramón y Cajal s/n, 28040 Madrid, Spain

[3.] CIBER de Bioingeniería, Biomateriales y Nanomedicina (CIBER-BBN), Spain

(\*)  Corresponding authors:

M. Vallet Regi (vallet@ucm.es), A. R. Boccaccini (aldo.boccaccini@ww.uni-erlangen.de)

**Abstract**

Ion-doped binary $SiO_2$-$CaO$ and ternary $SiO_2$-$CaO$-$P_2O_5$ mesoporous bioactive glasses were synthesised and characterised to evaluate the influence of $P_2O_5$ in the glass network structure. Strontium, copper and cobalt oxides in a proportion of 0.8 mol% were selected as dopants because the osteogenic and angiogenic properties reported for these elements. Although the four glass compositions investigated presented analogous textural properties, TEM analysis revealed that the structure of those containing $P_2O_5$ exhibited an increased ordered mesoporosity. Furthermore, $^{29}Si$ NMR revealed that the incorporation of $P_2O_5$ increased the network connectivity and that this compound captured the $Sr^{2+}$, $Cu^{2+}$ and $Co^{2+}$ ions preventing them to behave as modifiers of the silica network. In addition, $^{31}P$ NMR results revealed that the nature of the cation directly influences the characteristics of the phosphate clusters. In this study, we have proven that phosphorous oxide entraps doping-metallic ions, granting these glasses with a greater mesopores order.

## 1. Introduction

Bioactive glasses (BGs) are known to have an outstanding capability to stimulate bone regeneration [1] by their bone-bonding abilities. Two processing methods are mainly used to prepare these glasses: the traditional melt-quenching process and the sol-gel technique. One of the main advantages of sol-gel glasses is their high bioactive behavior due to the rapid dissolution of this material originated from their fine porous structure and the important quantity of silanol groups present in their surface, which could act as nucleation sites for the formation of an apatite layer when in contact with simulated body fluid (SBF)[2–4]. Furthermore, in 2004[5] a new generation of bioactive glasses, namely mesoporous bioactive glasses (MBGs), was introduced. MBGs are characterized by presenting tailored (ordered) porosity at the nanometer scale, which is created by combination of supramolecular chemistry of surfactants and sol-gel chemistry and opened new application fields to these materials, such as drug delivery. Research on the application of MBGs in tissue engineering and drug delivery has been accelerated in recent years as emphasized in recent reviews[6–8]. These materials, whose composition is mostly based on the $SiO_2$-$CaO$-$P_2O_5$ system, have a highly ordered mesoporous structure, which induces an accelerated bioactive behavior and the capacity of confining drug molecules to be released in a controlled manner[6]. It has also been reported that the addition of phosphorous is not necessary to obtain silicate MBGs of high bioactivity [9]. However, $P_2O_5$ plays a significant role on the local structure of the MBGs network. In this sense, it has been reported that most of the phosphorous atoms are incorporated as orthophosphates within the silica network, thus resulting in the formation of small clusters with $Ca^{2+}$ cations [10] that highly determines the bioactive behavior of these glasses.

During the last decade, metallic ions such as boron, copper, lithium, gallium, silver, strontium, cerium, cobalt and zinc, among others, have emerged as potential therapeutic agents with the ability to enhance bone formation by their stimulating effects on osteogenesis and angiogenesis[11,12]. In recent studies, new compositions of mesoporous materials including the addition of therapeutic ions have been suggested in order to increase the materials' functionality and biological activity, particularly for applications in large-size bone defects, which are a significant clinical challenge[8]. A main property of some of the commonly used therapeutic ions such as boron, zinc, cerium and gallium, is that they are known to be osteogenic. Other functional properties such as angiogenesis and antibacterial activity can also be assigned to some therapeutic ions[8,11,12].

Notably, copper, strontium and cobalt ions have attracted particular interest as they have been reported to confer osteogenesis, angiogenesis and antibacterial properties to MBGs[8]. $Cu^{2+}$ ions are known to stimulate the proliferation of endothelial cells, enhance cell activity and proliferation of osteoblastic cells, improve micro-vessel formation as well as to promote wound healing and to have antibacterial effects[8,12,13]. Furthermore, copper ions are not affected during scaffolds processing which involves high temperatures[14]. $Sr^{2+}$ ions have shown to be able to enhance bone cell activities by promoting osteoblast activity and inhibiting osteoclast differentiation, as well as maintain excellent acellular bioactivity of the glasses[8,12]. Furthermore, it has been observed that the release of $Sr^{2+}$ ions also stimulates significantly ALP activity[8,12]. Nevertheless, doping amount within the glass must be carefully controlled as a high content of Sr will influence the glass network formation and decrease the degree of order of the mesoporous structure[8]. $Co^{2+}$ ions are described in the literature as having angiogenic capacity determined by inducing a hypoxic cascade (including HIF-1α stabilization and VEGF secretion from hBMSCs)[10,15]. Generally, $Co^{2+}$ ions are added to the composition in order to develop hypoxia-mimicking (low oxygen pressure environment) materials[8,12,15, 16]. It has been reported that by mimicking hypoxic condition, $Co^{2+}$ ions could induce the coupling of osteogenesis and angiogenesis[15]. Nonetheless, it has also been reported that $Co^{2+}$ ions lead easier to potential cytotoxicity compared to $Cu^{2+}$ ions[8], an important aspect to be considered when doping the material with this ion. However, the role that even small amounts of these dopants can play on the mesoporous structure of MBG remains unclear.

In the present study, we attempt to demonstrate that the presence of phosphorous oxide in the composition allows capturing selected ionic dopants in the network, namely Sr and Co, thus conferring the material porosity of greater order. The incorporation of SrO and CuO in MBGs in an independent way has been previously reported in literature [8]. In the present investigation, both dopants were incorporated with the aim of achieving a synergy role of both compounds for future biological assays. Results of structural characterization to elucidate the influence of the ion content on the developed MBGs are presented. Based on previous work performed on ion-doped glasses[17–20], the range of substituent's concentration was chosen so that they did not inhibit or decrease the bioactivity of the MBGs. Moreover, the amounts of the leached ions ($Cu^{2+}$, $Sr^{2+}$, $Co^{2+}$) were chosen to be below the toxic level in blood plasma[8,11,12,20].

## 2. Materials and methods

### 2.1 Preparation of ion-doped mesoporous glasses

Ion-doped mesoporous glasses in the systems 78$SiO_2$ 20CaO 1.2$P_2O_5$ 0.8$X_{ion\ oxide}$ and 79.2$SiO_2$ 20CaO 0.8$X_{ion\ oxide}$ (in mol%) were synthesized, keeping the total amount of ion constant. The glasses were doped on the one hand with a mixture of strontium and copper ions, on the other hand with cobalt ions. The MBGs were prepared by the evaporation-induced self-assembly method (EISA)[20,21] and using Pluronic® F127 as structure-directing agent. Pluronic® F127 was dissolved overnight in 85 mL EtOH with 1.2 mL $HNO_3$ (0.5N). Reactants were then added under continuous stirring in the following order: tetraethyl orthosilicate (TEOS), in the case of the phosphorous containing compositions triethyl phosphate (TEP), calcium nitrate $Ca(NO_3)_2·4H_2O$ and strontium nitrate $Sr(NO_3)_2$ or copper nitrate $Cu(NO_3)_2·2.5H_2O$ or cobalt nitrate $Co(NO_3)_2·6H_2O$ (Sigma-Aldrich, Germany), with a 3 hours interval between each addition. The reaction was left under constant stirring at room temperature (RT) for 24 hours, and was then poured into petri-dishes (15 mL/petri-dish) for EISA during 5 to 7 days before proceeding with the heat treatment. The dried gels were calcinated at 700°C for 3 hours. The exact amounts of reactants are mentioned in Table 1.

Table 1 Amount of reactant for the synthesis of the different developed ion-doped glass compositions.

|  | TEOS [mL] | TEP [mL] | Ca(NO$_3$)$_2$ [g] | Sr(NO$_3$)$_2$ [g] | Cu(NO$_3$)$_2$ [g] | Co(NO$_3$)$_2$ [g] | F127 [g] |
|---|---|---|---|---|---|---|---|
| **SrCu/ P$_2$O$_5$** | 9.597 | 0.112 | 2.576 | 0.046 | 0.05 | - | 4.5 |
| **Co/ P$_2$O$_5$** | 9.597 | 0.112 | 2.576 | - | - | 0.128 | 4.5 |
| **SrCu/no P$_2$O$_5$** | 9.74 | - | 2.576 | 0.046 | 0.05 | - | 4.5 |
| **Co/no P$_2$O$_5$** | 9.74 | - | 2.576 | - | - | 0.128 | 4.5 |

## 2.2 Characterization of the ion-doped MBGs

Powder X-ray diffraction (XRD) analysis was performed using a D8 ADVANCE diffractometer (Bruker, Madison, US) in a 2θ range from 20 -80°. BG powders were dispersed in ethanol. Then, the solution was dripped on off-axis cut, low background silicon wafers (Bruker AXS, Germany). For the small angle measurements Philips X'Pert diffractometer equipped with Cu K$_α$ radiation (wavelength 1.5418 Å) was used. XRD patterns were collected in the 2θ° range between 0.5 θ° and 6.5 θ°, with a step size of 0.02 θ° and counting time of 4s per step.

Scanning electron microscope (SEM) micrographs of samples were recorded with an Auriga SEM Instrument (Zeiss, Germany). Energy dispersive spectroscopy (EDS) (Oxford Instruments) was performed to obtain a qualitative appreciation of the ions in the composition.

For transmission electron microscopy (TEM) analysis, samples were dispersed with ethanol on a lacey carbon film. In order to observe the mesoporous structure of the materials a JEOL-3010 transmission electron microscope operating at an acceleration voltage of 300 kV was used. Images were obtained using a CCD camera (model Keen view, SIS analyses size 1024 x 1024, pixel size 23.5 mm x 23.5 mm) at 30 000x and 60 000x magnification using a low-dose condition. Fourier transform (FT) patterns were conducted using a digital micrograph (Gatan).

Nitrogen physisorption isotherms were measured at 77 K using Micrometrics ASAP 2010 and ASAP 2020 porosimetry instruments. Prior to the measurements the samples were outgassed at least for 12 h at 120 – 150°C under vacuum. Pore size distribution curves were calculated from the adsorption branch of the isotherm according to the Barrett, Joyner and Halenda (BJH) method to determine mesoporous size distribution. The total pore volume was calculated from the adsorption point at 0.98 P/P$_0$. Specific surface areas were obtained from the BET equation in the linear range between 0.07 and 0.23 P/P$_0$.

$^1$H→$^{29}$Si and $^1$H→$^{31}$P CP (cross-polarization)/MAS (magic-angle-spinning) and single-pulse (SP) solid-state nuclear magnetic resonance (NMR) measurements were performed to evaluate the different silicon and phosphorus environments in the synthesized samples. The NMR spectra were recorded on a Bruker Model Avance 400 spectrometer. Samples were spun at 10 kHz for $^{29}$Si and 6 kHz in the case of $^{31}$P. Spectrometer frequencies were set to 79.49 and 161.97 MHz for $^{29}$Si and $^{31}$P,

respectively. Chemical shift values were referenced to tetramethylsilane (TMS) and $H_3PO_4$ for $^{29}$Si and $^{31}$P, respectively. All spectra were obtained using a proton enhanced CP method, using a contact time of 1 ms. The time period between successive accumulations was 5 and 4 s for $^{29}$Si and $^{31}$P, respectively, and the number of scans was 10 000 for all spectra.

3. **Results and discussion**

The developed samples $SrCu/P_2O_5$ and $Co/P_2O_5$ were heat treated at 700°C. In order to confirm the amorphous state of the samples, XRD analyses were conducted. Results are shown in Figure 1.

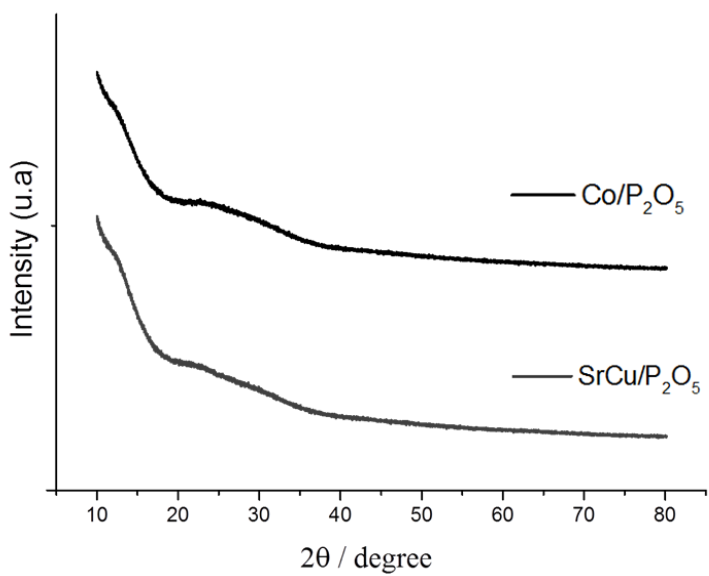

**Figure 1.** High angle XRD spectra of samples $SrCu/P_2O_5$ and $Co/P_2O_5$.

No diffraction peaks are observed from the XRD spectra (Figure 1) of samples $SrCu/P_2O_5$ and $Co/P_2O_5$ indicating the amorphous state. A broad reflection in the range between $2\theta$= 20 and 35 ° is clearly noticed for both samples suggesting that no crystalline phase is developed within the sample. Similar curves were obtained for $SrCu/noP_2O_5$ and $Co/noP_2O_5$ samples (not shown).

The prepared samples were doped with strontium and copper ions for samples $SrCu/P_2O_5$ and $SrCu/no\ P_2O_5$ and cobalt ions for samples $Co/P_2O_5$ and $Co/noP_2O_5$. Figure 2 displays the general aspect of the two different ion-doped systems. The particular colors of the samples come directly from copper for samples $SrCu/P_2O_5$ and $SrCu/no$ and cobalt for samples $Co/P_2O_5$ and $Co/noP_2O_5$.

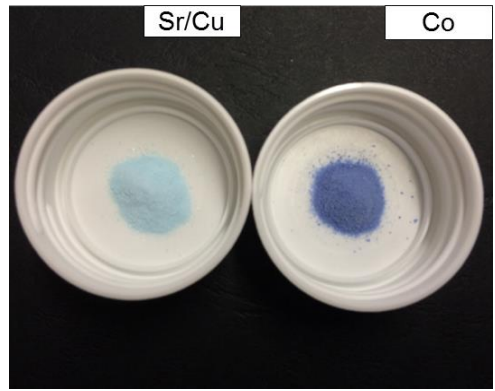

**Figure 2.** Image showing optical characteristics of samples doped with SrCu ions (on the left) and with Co ions.

A first appreciation of the homogeneity of the doped glasses was revealed by EDX analysis, as shown in Figure 3.

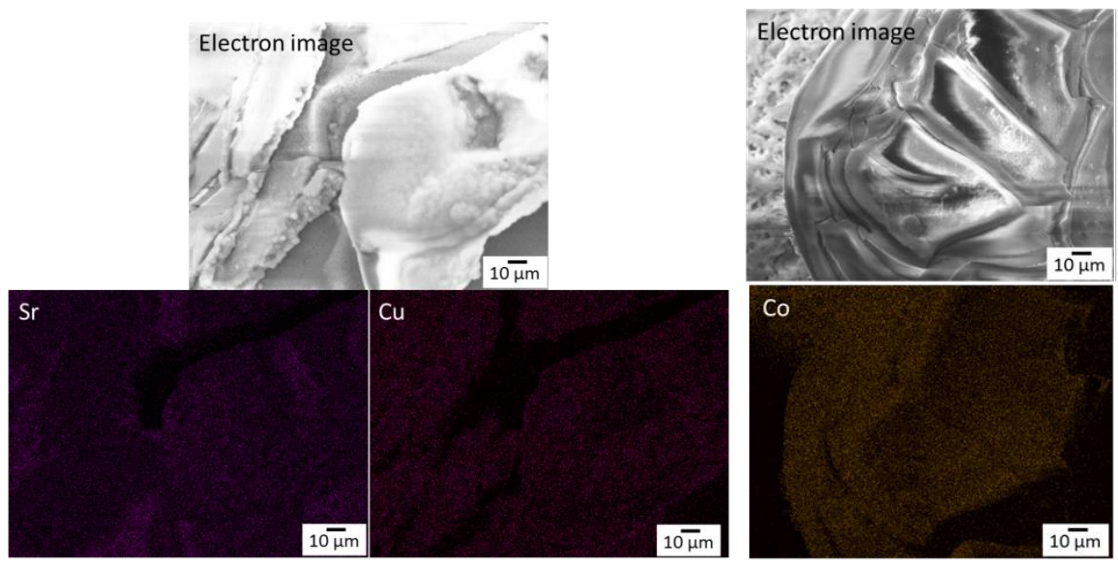

**Figure 3.** Images showing SEM and EDX mapping analysis (Sr, Cu and Co elements) for sample SrCu/$P_2O_5$ (left side) and sample Co/$P_2O_5$ (right side).

From images of Figure 3, it is observed that ions are well dispersed on the surface of the prepared materials. We have to take into consideration that EDS analysis is purely qualitative, thus the intensity of the elements coloured in the mapping images is not a quantitative representation of the doped-element amounts. Samples SrCu/no$P_2O_5$ and Co/no$P_2O_5$ showed similar results.

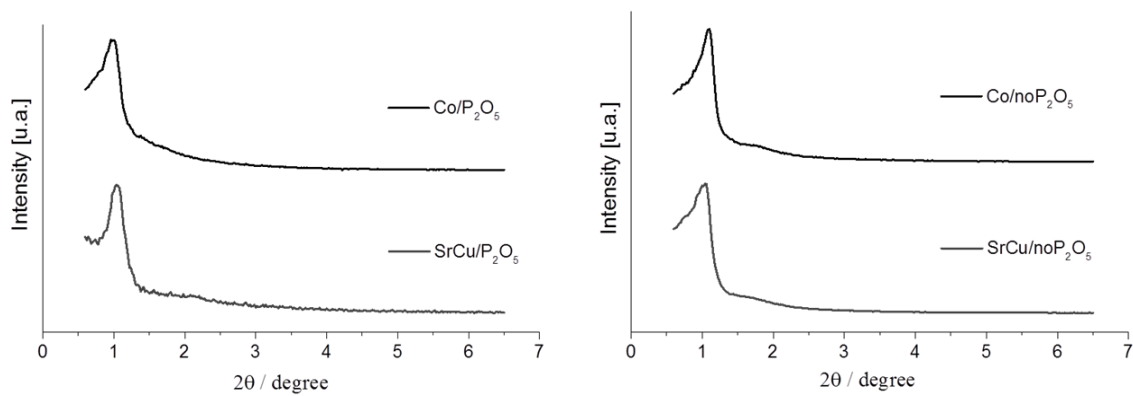

**Figure 4** SAXRD spectra of SrCu/P$_2$O$_5$ and Co/P$_2$O$_5$ (left side) as well as SrCu/noP$_2$O$_5$ and Co/noP$_2$O$_5$ (right side) measured in the small angle range (2θ: 0.5 to 8).

Figure 4 shows the SAXRD spectra of the four samples, namely SrCu/P$_2$O$_5$, SrCu/noP$_2$O$_5$, Co/P$_2$O$_5$ and Co/noP$_2$O$_5$. The four patterns exhibit a unique peak at around 2θ = 1.20°, typical for mesoporous materials, which could be attributed to the (1 0) maxima of a 2D-hexagonal p6m phase [4,17,22–24]. However, the information derived from these XRD patterns is not enough to assign a specific ordered phase for each sample and further evidences, as those provided by TEM, are required.

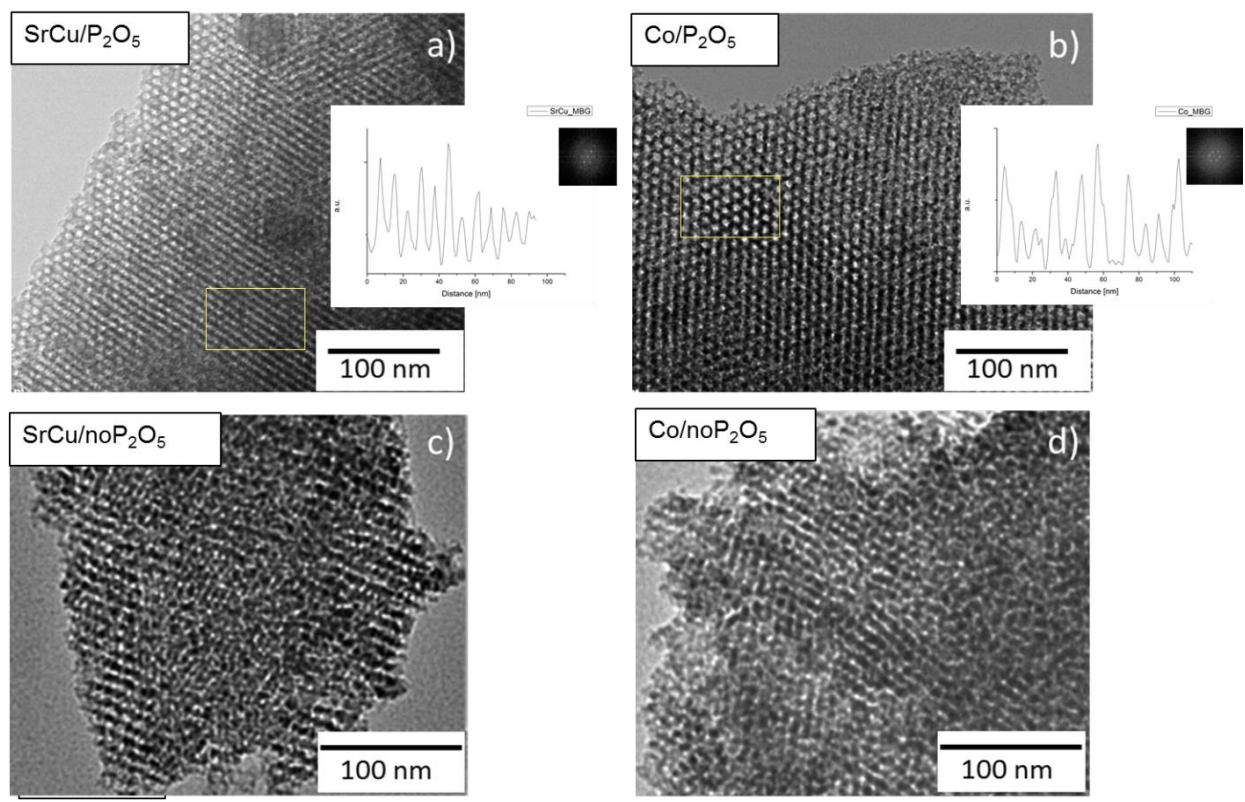

**Figure 5** TEM images showing the mesoporous structure of a) sample SrCu/P$_2$O$_5$, b) sample Co/P$_2$O$_5$, c) sample SrCu/noP$_2$O$_5$ and d) sample Co/noP$_2$O$_5$.

Figure 5 revealed results obtained by TEM studies. After carrying out the observations of at least 6 crystals of each sample, it was concluded that a clear ordered 2D-hexagonal mesoporous structure is observed for all $P_2O_5$ containing samples. The images shown in figure 5 are only the representative ones of a much wider study. Visibly, sample Co/$P_2O_5$ appears to have a lightly higher ordered mesoporosity than sample SrCu/$P_2O_5$, which is likely due to the double element doping for sample SrCu/$P_2O_5$ that tends to decrease the order in the structure. For samples prepared without TEP addition during the synthesis, a clear decrease in the order of the structure is observed. Samples SrCu/no$P_2O_5$ and Co/no$P_2O_5$ exhibit porosity in the mesosize, but the order is only kept in small-range regions, whereas most of the material exhibits a very defective structure, which is difficult to assign to a specific space group.

$N_2$ physical sorption analysis were performed in order to determine pore-related parameters, such as surface properties, specific surface area, pore size distribution and pore volume.

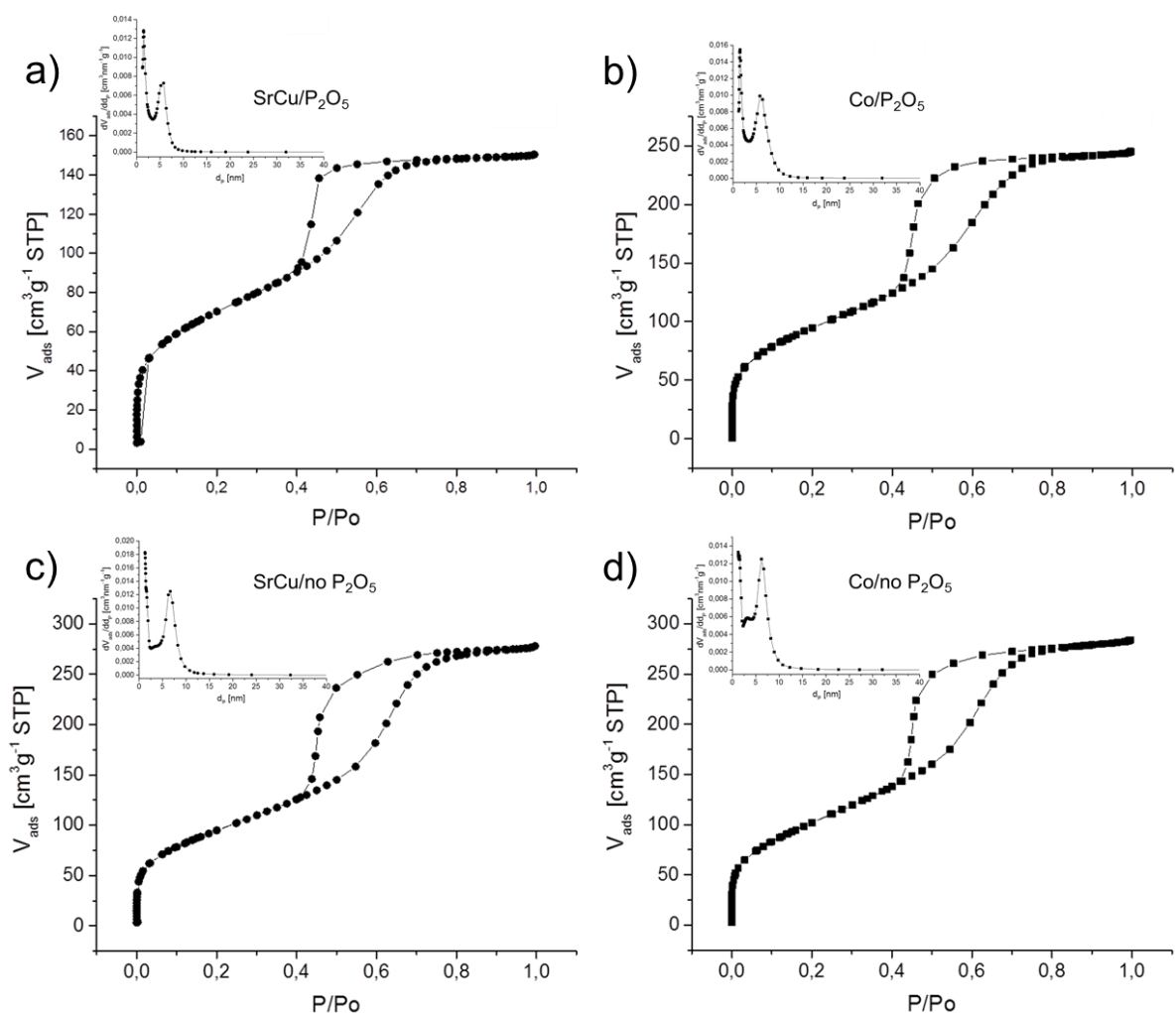

**Figure 6**. Nitrogen adsorption isotherms of the synthesized samples and pore size distribution calculated from the adsorption isotherms (insets).

Nitrogen adsorption isotherms of the four samples are shown in Figure 6. Table 2 lists the data of the surface area, pore volume and pore size of samples SrCu/$P_2O_5$, SrCu/no$P_2O_5$, Co/$P_2O_5$ and Co/no$P_2O_5$. The data presented for the four prepared samples is typical of mesoporous structures. The isotherm obtained for all samples, SrCu/$P_2O_5$, Co/$P_2O_5$, SrCu/no$P_2O_5$ and Co/no$P_2O_5$, are identified as type IV, which is the general case for mesoporous materials presented in the literature[19]. The sharp curve in the capillary condensation region is an indicator of narrow pore-size distribution. It has been reported[25] that hysteresis loops are included in isotherms obtained from those mesoporous materials with pore sizes large enough to allow the $N_2$ condensation within them. Hysteresis loop shown to differ among samples. In the case of SrCu/no$P_2O_5$ sample the loop exhibits slope changes in both adsorption and desorption branches, indicating a defective mesoporous structure. When $P_2O_5$ is incorporated (sample SrCu/$P_2O_5$), the loop can be identified as type H2, corresponding to ink-bottle pores[4,23], that is, pores narrow along their morphology that could be a consequence of the formation of phosphate based clusters with the different cations incorporated. Besides, the sharp slopes in both branches seem to indicate a higher ordering after $P_2O_5$ incorporation, in agreement with the results observed by TEM. The isotherms obtained for Co/$P_2O_5$ and Co/no$P_2O_5$ samples exhibit a similar situation, although in this case the loop of Co/no$P_2O_5$ seems to be less defective in comparison with SrCu/nO$P_2O_5$. Pore size distribution curves were calculated from the adsorption branch using the BJH model, which is also suited for this kind of mesoporous material. Samples SrCu/$P_2O_5$ and Co/$P_2O_5$ present a single-modal pore size distribution of. The narrow size distribution is centered at 3 nm for sample SrCu/$P_2O_5$ and at 3.6 nm for sample Co/$P_2O_5$. The BET surface areas of samples SrCu/$P_2O_5$ and Co/$P_2O_5$ reached 254 and 346 $m^2$/g respectively and the total pore volumes calculated at $P/P_0 = 0.99$ were 0.233 $cm^3$/g for sample SrCu/$P_2O_5$ and 0.379 $cm^3$/g for sample Co/$P_2O_5$. These results are in accordance with the literature, where mesoporous glasses are reported to exhibit $S_{BET}$ in the range of 200 - 500 $m^2$/g depending on the silica content[7]. Samples prepared without the addition of TEP during the synthesis do not show any difference in structural characteristics, thus this stage it is not possible to evaluate the advantage of adding phosphorous oxide. Besides, the pore size distribution for samples without $P_2O_5$ seems to be wider and centered in larger pore diameter, indicating a pore size heterogeneity associated with defective mesoporous structures. The role of $P_2O_5$ has been widely studied in MBGs by HRTEM and NMR. Studies have shown that this oxide induces the formation of orthophosphate nanoclusters, which can produce changes in the local environment of MBGs. These changes could be responsible for the apparent anomalies found in the pore size distribution.

**Table 2.** Structural parameters of samples SrCu/$P_2O_5$, SrCu/no$P_2O_5$ and D, Co/no$P_2O_5$

| Samples | $S_{BET}$ [$m^2$/g] | $V_p$ [$cm^3$/g] | $d_{p,mean}$ [nm] |
|---|---|---|---|
| SrCu/$P_2O_5$ | 254 | 0.233 | 3.0 |
| Co/$P_2O_5$ | 346 | 0.379 | 3.6 |
| SrCu/noP$_2$O$_5$ | 346 | 0.417 | 6.6 |
| Co/noP$_2$O$_5$ | 377 | 0.439 | 6.4 |

In our study, TEM and porosimetry studies indicate that the incorporation of P$_2$O$_5$ tends to increase ordered mesoporosity (samples SrCu/P$_2$O$_5$ and Co/P$_2$O$_5$ compared to SrCu/noP$_2$O$_5$ and Co/noP$_2$O$_5$ without addition of TEP). It is suggested that the addition of this component in the doped-glass allows incorporating more efficiently the cations within the network by cluster formation. In order to corroborate this hypothesis solid state NMR analysis was carried out to investigate the details of the network structure.

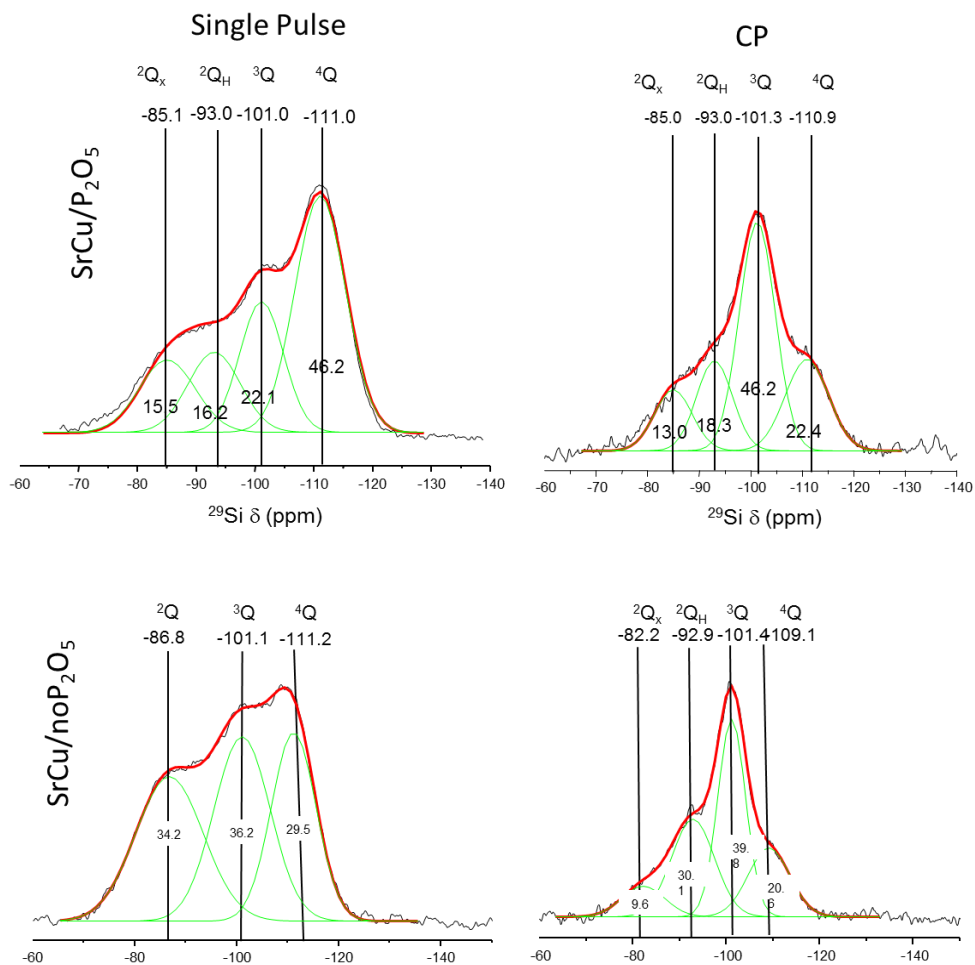

**Figure 7.** Solid-state $^{29}$Si single pulse (left) and $^1$H cross-polarization (right) MAS NMR spectra (red) for solids SrCu/P$_2$O$_5$, SrCu/noP$_2$O$_5$. The areas for the Q$^n$ units were calculated by Gaussian line-shape deconvolutions (green).

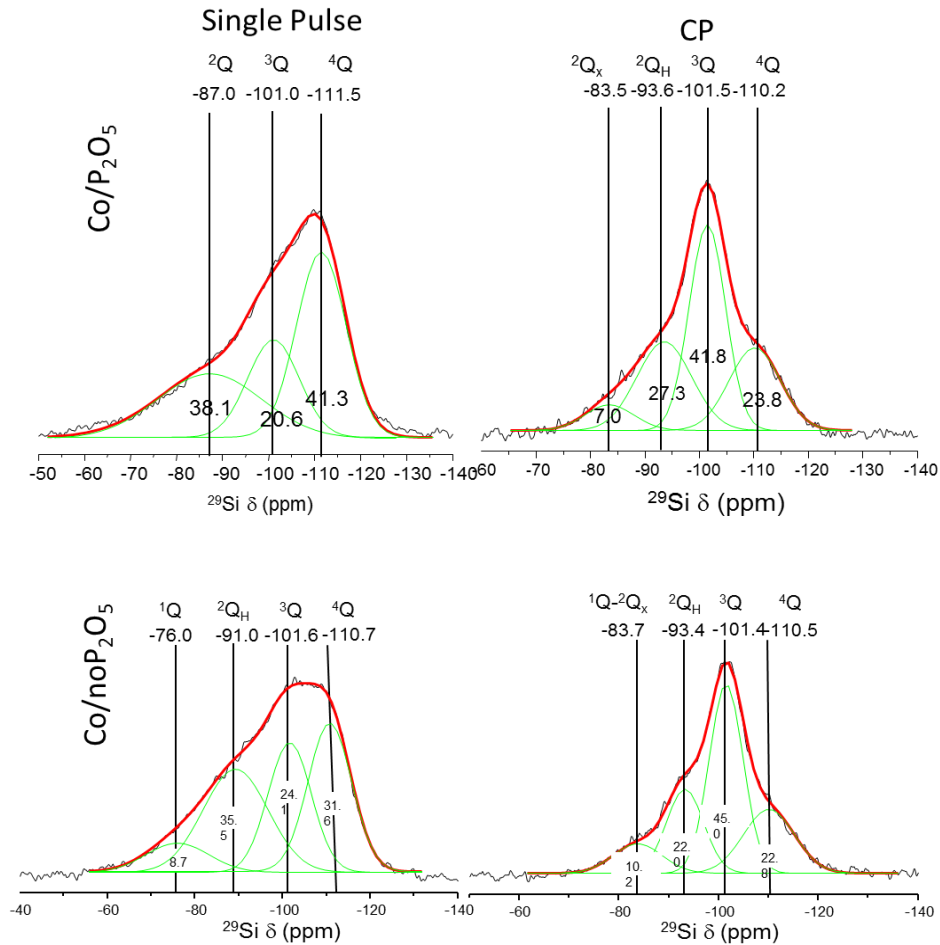

**Figure 8** Solid-state $^{29}$Si single pulse (left) and $^1$H cross-polarization (right) MAS NMR spectra (red) for solids Co/P$_2$O$_5$, Co/noP$_2$O$_5$. The areas for the Q$^n$ units were calculated by Gaussian line-shape deconvolutions (green).

$^{29}$Si NMR spectroscopy was used to evaluate the network connectivity of MBGs as a function of the presence or absence of phosphorous in the chemical composition. Q$^1$, Q$^2$, Q$^3$ and Q$^4$ represent the silicon atoms (denoted Si*) in (NBO)$_3$Si*-(OSi)$_1$ (NBO)$_2$Si*-(OSi)$_2$, (NBO)Si*-(OSi)$_3$, and Si*(OSi)$_4$ (NBO = non-bonding oxygen), respectively. The signals in the region of -109 ppm to -112 ppm are from Q$^4$, those in the region -100 ppm to -102 ppm are from Q$^3$. The Q$^2$ signal appears approximately at -93 ppm to -85 ppm. Q$^2$ signals can be divided into two different signals associated with different types of silicon, related to with hydrogen and associated with network modifiers. On the one hand, Q$^2_H$ signals appear at -91 to -93 ppm and Q$^2_x$ is associated with signals between -82 to -85 ppm. Finally, signals at lower shifts (~76 ppm) are assigned to Q$^1$ units. [10, 27]

Figure 7 and Figure 8 show Single Pulse $^{29}$Si (left) and Cross Polarization (right) spectra for SrCu and Co doped samples, respectively. Single pulse spectra show higher contribution of Q$^4$ compared to $^1$H CP spectra, as could be expected from the lower connectivity of silica tetrahedron in those regions enriched in protons, i.e. the glass surface. On the contrary, the silica units located at the surface exhibit higher amounts of Q$^3$ units that would correspond with the presence of Si-OH groups. Table 3 shows the chemical shifts, deconvoluted peak areas, and silica network connectivity of the four samples studied in this work. By comparing samples with and without $P_2O_5$, it is clear that the incorporation of this compound leads to an increase of network connectivity. This fact evidences that $P_2O_5$ entraps the divalent cations and removes them from the silica network.

**Table 3** Chemical shifts, relative peak areas, and silica connectivity obtained by Solid-state single pulse $^{29}Si$ MAS NMR

| | $Q^4$ | | $Q^3$ | | $Q_H^2$ | | $Q_X^2$ | | |
|---|---|---|---|---|---|---|---|---|---|
| | $\delta_{Si}$ (ppm) | % | $\delta_{Si}$ (ppm) | % | $\delta_{Si}$ (ppm) | % | $\delta_{Si}$ (ppm) | % | NC |
| SrCu/$P_2O_5$ | -111.0 | 46.2 | -101.0 | 22.1 | -93.0 | 16.2 | -85.1 | 15.5 | 3.13 |
| SrCu/no$P_2O_5$ | -111.2 | 29.5 | -101.1 | 36.2 | -86.8 | 34.2 | - | - | 2.95 |
| Co/$P_2O_5$ | -111.5 | 41.3 | -101.0 | 20.6 | -87.0 | 38.1 | - | - | 3.02 |
| Co/no$P_2O_5$ | -110.7 | 30.3 | -101.6 | 52.2 | -91.0 | 35.5 | -76.0 | 8.7 | 2.78 |

$^{31}$P-NMR results were used to evaluate the local environment of P atoms, thus elucidating the phosphate species contained in the different samples (Figure 9). The spectra recorded by single pulse show one signal at ~2 ppm assignable to q0 units present in the $PO_4^{3-}$ species. The signals for Co/$P_2O_5$ show full width at half-maximum height (FWHM) of around 12 ppm, which is typical of an amorphous orthophosphate[28]. On the contrary, the FWHM for SrCu/$P_2O_5$ sample is much narrow indicating that the clusters are larger and more crystalline in this sample. These results evidence that most of the P atoms are included as independent $PO_4^{3-}$ tetrahedra within the silica network, thus avoiding polyphosphate formation. The orthophosphates would be balanced with the divalent cations introduced in the system and the nature of the cations ($Sr^{2+}$, $Cu^{2+}$ or $Co^{2+}$) seems to play an important role on the characteristics of the phosphate clusters.

It has been reported in the literature[8] that MBGs are efficient systems for the simultaneous delivery of drugs and therapeutic ions. These delivery properties confer multifunctional features to mesoporous glasses which may enhance considerably the regeneration of large-bone defects. Although the beneficial effects of these cations have been widely reported, our results point out to an important drawback associated to their incorporation, i.e. the mesoporous structure is significantly affected even with small amounts of these dopants. This effect could be related with the high polarization effects of $Co^{2+}$ and $Cu^{2+}$, or with the large size of $Sr^{2+}$ cations. In previous studies[25], the capability of phosphorous oxide entrapping divalent cations such as $Ca^{2+}$ has been demonstrated. The consequence is that cations are removed from the silica network, thus suppressing their role as network modifiers and leading to higher silica network connectivity.

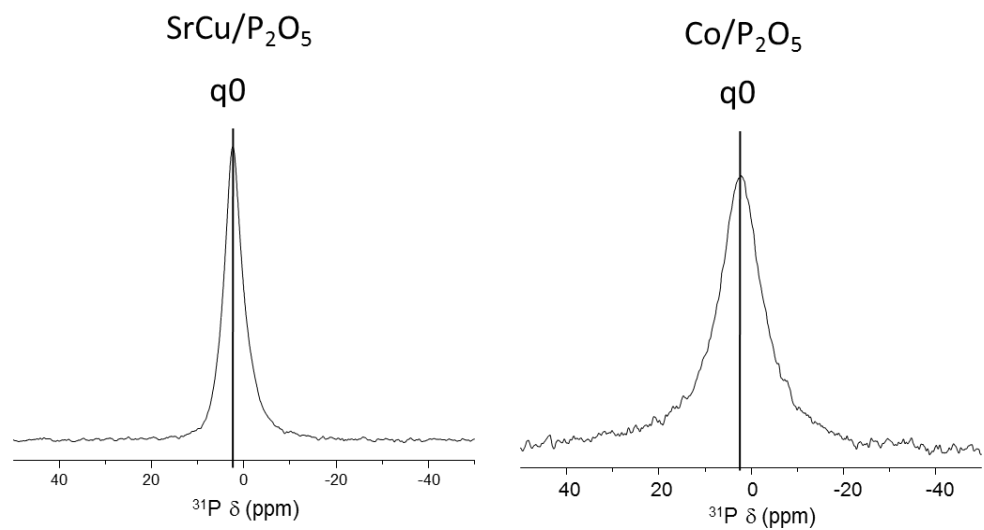

**Figure 9** Solid-state $^{31}$P single-pulse MAS NMR spectra of SrCu/P$_2$O$_5$ and Co/P$_2$O$_5$ samples.

## 4. Conclusions

Mesoporous bioactive glasses doped with Sr$^{2+}$/Cu$^{2+}$ or Co$^{2+}$ have been synthesized. In order to evaluate the role of P$_2$O$_5$ in the mesoporous structure, two series with and without P$_2$O$_5$ were prepared. The incorporation of a small amount (0.8 mol%) of Sr$^{2+}$/Cu$^{2+}$ or Co$^{2+}$ ions was shown to lead to defective mesoporous structures, as evidenced by TEM. However, when 1.2 mol% of P$_2$O$_5$ is present, the MBGs exhibited a highly ordered mesoporous structure. This fact is due to the capability of P$_2$O$_5$ to form clusters with divalent cations, thus avoiding these dopants to interfere during the mesophase formation. The findings of this work indicate an effective way to keep the high mesoporous structure of MBGs, even when cations such as Sr$^{2+}$/Cu$^{2+}$ or Co$^{2+}$ are incorporated during the synthesis to produce multifunctional MBGs.

**Acknowledgments**

This study was supported by research grants from the Ministerio de Economía y Competitividad (projects MAT2013-43299-R and MAT2015-64831-R), European Research Council (ERC-2015-AdG). Advanced Grant Verdi-694160, Agening Network of Excellence (CSO2010-11384-E), Instituto de Salud Carlos III (PI15/00978), and EU-ITN project GlaCERCo (GA 264526). The authors thank the staff of the ICTS National Center for Electron Microscopy, UCM, Madrid (Spain) for the assistance in the scanning and transmission electron microscopy.